\documentstyle[12pt]{article}

\begin{document}
\begin{titlepage}
\begin{center}
\today \hfill OHSTPY-HEP-T-94-023 \\
\hfill   LBL-36357,\\
\hfill   UCB-PTH-94/27 \\
\vskip .25in

{\large\bf A Complete Supersymmetric SO(10) Model}

 \vskip .25in

{{Lawrence J. Hall} \footnote{Supported in part by U.S. Department of
Energy contract DE- AC03-76SF00098, and by the National Science Foundation
under grant PHY-90-21139} \\
{\it Department of Physics, University of
California at Berkeley}\\ {\it and}\\
      {\it Theoretical Physics Group, Lawrence Berkeley Laboratory}\\
     {\it 1 Cyclotron Road, Berkeley, California 94720}\\
{\it hall\_lj@theorm.lbl.gov} \\ {and}
\\ {Stuart Raby} \footnote{Supported in part by U.S. Department of Energy
contract DE-ER-01545-640.}\\{\it Department of Physics, The Ohio State
University}  \\{\it 174 W. 18th Ave., Columbus, OH 43210}\\
{\it raby@mps.ohio-state.edu}}
\end{center}

\vskip .5in

\begin{abstract}

A complete supersymmetric SO(10) model is constructed, which is the most
general consistent with $R$, discrete, and $U(1)$ flavor symmetries.  At the
supersymmetric level there are many degenerate vacua, one of which is
phenomenologically successful.
This vacuum has vevs which align in certain definite directions in SO(10)
group space, such as the $B-L$ direction. Although this desired vacuum is not
proven to be the global minimum of the entire theory, including supersymmetry
breaking, it is separated from vacua where the vevs point in different group
directions by a large potential barrier. This desired vacuum simultaneously
leads to three major features of the theory.  (1) SO(10) is broken at scale
$v_{10}$ to SU(5), which breaks at $M_G$ to the standard model gauge group.
Beneath $M_G$ the only light gauge non-singlet fields are those of the minimal
supersymmetric standard model, so that the successful prediction for the weak
mixing angle is retained. (2)  The alignment of the vevs leads to a natural
mass separation of the weak Higgs doublets from their colored partners
via a mechanism which is closely related to the issue of the proton decay rate.
Also, the generation of the $\mu$ term is studied.  (3)  A set
of particles acquire mass at the highest perturbative scale of the theory, $M$
and at $v_{10}$.  When they are integrated out, they lead to just four flavor
operators for quark and charged lepton masses, and two more for neutrino
masses. These flavor operators lead to many quark and
lepton mass and mixing angle relations which involve pure SO(10) group theory
numerical Clebsch factors. The family hierarchies result from the ratio of
scales $v_{10}/M, M_G/v_{10}$ and $M_G/M$. While the theory is by no means
unique, it is complete, and hence illustrates the close connecton between
several important features of the theory.

\end{abstract}
\end{titlepage}
\renewcommand{\thepage}{\roman{page}}
\setcounter{page}{2}
\mbox{ }

\vskip 1in

\begin{center}
{\bf Disclaimer}
\end{center}

\vskip .2in

\begin{scriptsize}
\begin{quotation}
This document was prepared as an account of work sponsored by the United
States Government. While this document is believed to contain correct
 information, neither the United States Government nor any agency
thereof, nor The Regents of the University of California, nor any of their
employees, makes any warranty, express or implied, or assumes any legal
liability or responsibility for the accuracy, completeness, or usefulness
of any information, apparatus, product, or process disclosed, or represents
that its use would not infringe privately owned rights.  Reference herein
to any specific commercial products process, or service by its trade name,
trademark, manufacturer, or otherwise, does not necessarily constitute or
imply its endorsement, recommendation, or favoring by the United States
Government or any agency thereof, or The Regents of the University of
California.  The views and opinions of authors expressed herein do not
necessarily state or reflect those of the United States Government or any
agency thereof or The Regents of the University of California and shall
not be used for advertising or product endorsement purposes.
\end{quotation}
\end{scriptsize}

\vskip 2in

\begin{center}
\begin{small}
{\it Lawrence Berkeley Laboratory is an equal opportunity employer.}
\end{small}
\end{center}

\newpage
\renewcommand{\thepage}{\arabic{page}}
\setcounter{page}{1}

\section{Introduction}

The Standard Model is a great success. In the coming years experiments will
continue to look for deviations from the Standard Model which might indicate
new physics.  There are already several suggestive hints of new physics.
\begin{itemize}
\item The measured values of $\alpha,  \sin^2\theta_W$ and $\alpha_s(M_Z)$ are
consistent with the prediction of a simple supersymmetric[SUSY] grand unified
theory[GUT] (with $\alpha_i \approx \alpha_G$, i = 1,2,3, at
$M_G \sim 10^{16} GeV $ and a soft SUSY breaking scale of order
$(1 - 10)\times M_Z$) \cite{DRW}.
\item  The measured values of $m_b, m_{\tau}$ and $m_t$  are consistent with
SUSY GUTs (with the Yukawa couplings satisfying the SU(5) relation
$\lambda_b = \lambda_{\tau}$ at $M_G$).  This constraint correlates $m_t$ and
the ratio of Higgs vacuum expectation values[vevs], $\tan \beta$.
For $m_t \sim 174 GeV$,  there is both a small and large solution for
$\tan \beta$. In fact, the SO(10) relation
$\lambda_b = \lambda_{\tau} = \lambda_t$ at $M_G$ is also consistent
with observation \cite{ALS}.  This constraint favors the solution with
$\tan \beta$ in the range 40 - 60.
\item The cosmological evidence for a universe filled with cold and hot dark
matter fits nicely into the minimal supersymmetric standard model[MSSM] with a
conserved R-parity. In this case the lightest SUSY particle is absolutely
stable and is an excellent candidate for the cold dark matter constituent. A
tau neutrino with mass of order several eV is the natural candidate for the
hot dark matter.  Such a tau neutrino may be observable at the CHORUS and
NOMAD experiments at CERN and at the E803 experiment at Fermilab.
\item Cosmological and astrophysical evidence for neutrino masses and mixing
angles is indicative of right-handed neutrino components which naturally fit
into an SO(10) framework for fermions.\footnote{Recently it has been shown
that acceptable values for $m_b/m_{\tau}$ together with a
right-handed neutrino Majorana mass of order $10^{12} GeV$
(as indicated, for example, by a tau neutrino component of dark matter)
requires large $\tan\beta$ \cite{BMR}.}
\end{itemize}

All of these hints are quite tantalizing.  As a whole they strongly suggest
the notion of a SUSY GUT.  It is an intriguing question then whether the
symmetry relations of a SUSY GUT can help us understand the observed pattern
of fermion masses and mixing angles. Recently, several effective SO(10) SUSY
GUT models were found which provide a consistent and quantitative description
of this low energy data \cite{ADHRS} (from now on referred to as paper I)
\cite{HISTORY}.

SO(10) \cite{SO(10)} has the advantage
that it is the minimal GUT in which all the fermions in
one family, i.e.  u, d, e and  $\nu$, fit into one irreducible representation
with only one additional state -- a right-handed neutrino. Thus mass matrices
in different charge sectors can be related. In reference I a search was made
for all acceptable SO(10) flavor sectors in which all
quark and charged lepton masses and mixings
originated in just four operators. Several such models of flavor were found to
be consistent with data.
This agreement may of course be completely spurious,
or these operators may give the dominant
contributions to fermion masses in the effective theory at $M_G$. If the latter
is true then corrections to the leading order results should improve this
agreement. Why should only a few operators contribute to fermion masses? Why
should nature be kind and allow us to make many flavor predictions
without addressing physics at the Planck scale? This is an important criticism
of reference I. The answer given in this paper is that flavor symmetries of
the theory just beneath the Planck scale can be very powerful, forbidding the
vast majority of possible operaters.

The main purpose of this paper is to provide a framework in which such
corrections may be calculated. There are several sources of corrections:
electroweak scale threshold corrections must be included, which can be
significant for large $\tan\beta$. GUT scale threshold corrections and GUT
scale corrections to the effective
theory can affect fermion masses as well as FCNC processes.  In addition, new
phenomena such as proton decay and neutrino masses can
only be addressed in the context of a complete SO(10) SUSY GUT valid for
energies greater than $M_G$
up to the largest perturbative scale in the theory, $M$, which can be taken as
the string compactification scale or the Planck mass.

In this paper we demonstrate the existence of a complete SO(10) SUSY GUT which
reproduces, as an example, one of the models found in I.  By a complete GUT we
mean one which reproduces not only a realistic fermion mass spectrum (including
neutrinos) but also addresses the issues of GUT symmetry breaking, the
doublet-triplet splitting problem, the $\mu$ problem and proton
decay.  We show how all these problems have reasonable solutions, with one
caveat.  We do not claim to understand the dynamics which determines
the scale of the GUT vevs.
All GUT breaking occurs along flat directions in the SUSY limit.
We will also not address the question of the origin of soft SUSY breaking.
We will just assume the standard set of soft SUSY breaking terms below M
whenever they are needed to make contact with low energy data.

We make no claim of uniqueness for the model in this paper: there are
undoubtedly many similar such models. Nevertheless, by studying a specific,
complete model, one can study the connections between many different aspects
of the theory. The orientation of the GUT vevs is crucial for understanding
both the lightness of the Higgs doublets and for obtaining the operators which
lead to predictions for the quark and lepton masses.
Similarly, the doublet-triplet
splitting problem is intimately connected with the proton decay rate. Finally,
a family symmetry, which is imposed to give some understanding of the observed
pattern of quark and lepton masses and mixings, has important consequences for
many other aspects of the theory.

In the next section we discuss the flavor sector of the effective theory at
$M_G$. It involves just four operators, and we study how this structure may
emerge from symmetries of the full theory. In sections 3 and 4 we focus on
the GUT symmetry breaking and Higgs potential respectively.  Section 4 also
includes the mechanisms for doublet-triplet splitting, for solving the
$\mu$ problem and constraints from proton decay.  In section 5 we consider
neutrino masses. Finally in section 6, we discuss the full set of symmetries
of the theory.  In section 7, we give our general conclusions and directions
for further work.

\section{Charged Fermion Masses}

We take the flavor sector of our model to be that of model 6 in paper I.
The fermion mass and mixing angle predictions which result from this flavor
sector, for a particular choice of input parameters, are shown in Table I,
reproduced from paper I.
\newpage
\begin{center}
\indent Table 1: Particular Predictions for Model 6
with $\alpha_s(M_Z) = 0.115$
\vskip 20pt
\begin{tabular}{|c|c|c|c|}
\hline
  Input  & Input & Predicted & Predicted  \cr
 Quantity &  Value &  Quantity & Value \cr
\hline
  $m_b(m_b)$ & $4.35 $ GeV & $M_t$ & $176$ GeV \cr
  $m_\tau(m_\tau)$ &$1.777 $ GeV & $\tan\beta$ & $55 $  \cr
\hline
  $m_c(m_c)$ &$1.22$ GeV & $V_{cb}$ & $.048 $ \cr
\hline
  $m_\mu$ &$105.6 $ MeV   & $V_{ub}/V_{cb}$ & $.059 $ \cr
  $m_e $   &$0.511$ MeV  &  $m_s(1GeV)$ & $172 $ MeV \cr
  $V_{us}$       &$0.221 $    & $\hat{B_K} $ & $0.64$  \cr
                 &            & $m_u / m_d$  & $0.64$  \cr
                 &            & $m_s / m_d$  & $24.$  \cr
\hline
\end{tabular}
\end{center}
In addition to these predictions, the set of inputs in
Table 1 predicts:
$\sin 2\alpha = -.46$, $\sin 2\beta = .49$,
$\sin 2\gamma = .84$, and $J = 2.6\times 10^{-5}$.

We have chosen model 6 of paper I since this model gives results which are in
best agreement with the data.  Model 9 of paper I generally gives
values of $V_{cb}$ which are larger and model 4 seems to have too
little CP violation. Of course these problems
may in fact be solved by corrections to these leading order results.

The flavor sector is specified by a particular set of four operators $\{
O_{33}, O_{32}$, $O_{22}, O_{12} \}$.  Three of these operators --
$O_{33}, O_{32}$, and $O_{12}$ -- are uniquely specified by choosing model 6.
On the other hand there
are 6 choices for operator $O_{22}$, as all give identical entries in the
charged fermion Yukawa matrices.
In order to construct the theory above $M_G$ one
of these operators must be chosen.  This ambiguity is real.  Perhaps further
study might reveal that only one of these operators is preferred by symmetry
arguments. In the meantime we have made a particular selection.

The four effective fermion mass operators chosen are given by
\begin{eqnarray}
    O_{33} = &  16_3\ 10_1 \ 16_3 & \\
     O_{23} = & 16_2 \ {A_2 \over {\tilde A}} \ 10_1 \ {A_2 \over
{\tilde A}} \ 16_3 &  \nonumber \\
    O_{22}  =  &  16_2 \ {{\tilde A} \over {\cal S}_M} \ 10_1 \ {A_1 \over
{\tilde A}} \ 16_2 & \nonumber \\
    O_{12} = & 16_1 \left( {{\tilde A}\over {\cal S}_M}\right)^3 \ 10_1 \left(
{{\tilde A} \over {\cal S}_M} \right)^3 16_2 & \nonumber
\end{eqnarray}
The adjoint fields $A_1, A_2, \tilde{A}$ are assumed to get vevs equal to
\begin{eqnarray}
 \langle A_1 \rangle  = & a_1 {3 \over 2} (B - L)  &  \\
 \langle A_2 \rangle = & a_2 {3 \over 2} Y  &  \nonumber \\
 \langle \tilde{A} \rangle = & - \tilde{a} X  &  \nonumber
\end{eqnarray}
respectively with $a_1, a_2 \sim M_G$ and $\tilde{a} \sim v_{10} > M_G$.  The
singlet ${\cal S}_M$ is assumed to get a vev $\sim M$.  The superspace
potential for these fields is discussed in the next section.

Consider the symmetries which are necessary in order to guarantee these and
only these fermion mass operators. First it is clear that adjoints with
distinct vevs appearing in the operators of eqn. 1 cannot be interchanged.
Interchanging them would lead to different Clebschs and a new, unacceptable
model.  For each distinct vev we need one adjoint chiral supermultiplet.
Also changing the position of the adjoints in each operator would alter the
Clebschs and again lead to an unacceptable theory. In order to prevent
interchangeability of the adjoints we assume they carry a different value
of one (or possibly more) U(1) charges.  However, we can only prevent the
positional changes of adjoints in these operators by studying the symmetries
of the theory above $v_{10}$, as we now demonstrate.

Each operator can be obtained via a unique tree diagram constructed with
dimension 4 couplings to intermediate heavy $16$ and $\overline{16}$ states.
In fig. 1 we show the explicit decomposition for the effective operators of
eqn. 1.

We now assign unique U(1) charges to the heavy $16, \overline{16}$ states such
that no  other dimension 4 couplings (other than those already appearing in
the vertices of fig. 1) are consistent with the charge assignment.
\underline{If}
this is possible then no other effective fermion mass operators can be
obtained to leading order when integrating out the heavy  $16, \overline{16}$
states.
Note the heavy $16, \overline{16}$ states have mass greater than $M_G$ since
the SO(10) singlet field ${\cal S}_M$ and adjoint $\tilde{A}$ are assumed to
have vevs of order $M, v_{10}$ respectively with $M \ge v_{10} > M_G$. The
choice for $O_{22}$ will affect the possible U(1) charges of the intermediate
states.  Not all choices
are allowed.  Note also that the Higgs 10 coupling to the fermion mass
operators carries the label 1.  Two 10s are necessary for the
doublet-triplet splitting mechanism used below,
but only  $10_1$ couples to fermions.  This will be enforced by symmetries.

The superspace potential for this sector is given by
\begin{eqnarray}
 W_{fermion} = &    &     \\
 &16_3 16_3 10_1  +  {\bar \Psi}_1 A_2 16_3  + & {\bar \Psi}_1 {\tilde
A} \Psi_1  +  \Psi_1 \Psi_2 10_1  \nonumber \\
 & +  {\bar \Psi}_2 {\tilde A} \Psi_2  +  {\bar \Psi}_2 A_2 16_2  +  &
{\bar \Psi}_3  A_1 16_2  \nonumber  \\
 & +  {\bar \Psi}_3 {\tilde A} \Psi_3  +  \Psi_3 \Psi_4 10_1  + & {\cal S}_M
\sum_{a=4}^9  ( {\bar \Psi}_a \Psi_a )  \nonumber \\
 & + {\bar \Psi}_4 {\tilde A} 16_2  +  {\bar \Psi}_5 {\tilde A} \Psi_4  + &
{\bar \Psi}_6 {\tilde A} \Psi_5  \nonumber  \\
 & +  \Psi_6 \Psi_7 10_1  +  {\bar \Psi}_7 {\tilde A} \Psi_8  + & {\bar
\Psi}_8  {\tilde A}
\Psi_9  +  {\bar \Psi}_9 {\tilde A} 16_1  \nonumber
\end{eqnarray}

The form of this superpotential is guaranteed by the symmetries discussed in
section 6. In particular, an $R$ symmetry forces it to be trilinear.

\section{GUT Symmetry Breaking Sector}

The superspace potential for the adjoints $A_1, A_2, \tilde{A}$  must preserve
their distinct U(1) charges.  We also need  $16$ and $\overline{16}$ fields
$\Psi$, $\overline{\Psi}$ to break the rank of SO(10) from  5 to 4.  Finally,
all states of this sector which are non-singlets under the standard model gauge
group  must get mass $\ge M_G$.
This is necessary so that the only states with mass $\le M_G$ are in the MSSM
or are singlets
which don't affect the RG equations from $M_G$ to $M_Z$. These constraints
are satisfied by the interactions
\begin{eqnarray}
W_{symmetry \,\, breaking} = &    &      \\
&   A'_1 (S A_1 + {\cal S}_1 A_1 ) +  S' ( {\cal S}_2 S + A_1^2) &
\nonumber \\ & + {\tilde A}' ( {\bar \Psi} \Psi + {\cal S}_3  \tilde A ) &
\nonumber  \\
 & +  A'_2 ( {\cal S}_4 A_2  +  S  {\tilde A} + ( {\cal S}_1 + {\cal S}_5)
 {\tilde A} ) & \nonumber \\
 & +   {\bar \Psi}' A_2 \Psi  +  {\bar \Psi} A_2 \Psi' & \nonumber \\
 &  +  A_1 A_2 {\tilde A}'  +  {\cal S}_6  (A'_1)^2 & \nonumber
\end{eqnarray}
where $S$ and $S'$ are 54 dimensional representations; $\Psi, \Psi'$ and
$\overline{\Psi}, \overline{\Psi}'$ are $16$ and $\overline{16}$ respectively
and $A_1, A_2, \tilde{A}, A_1', A_2', \tilde{A}'$ are adjoints. The primed
fields are necessary to preserve the distinct U(1) charge assignments for all
fields.  All primed fields are assumed to get zero vevs.

The superspace potential of eqn. 4 has many flat directions.  Once SUSY is
broken these flat directions will be lifted.  In this paper we shall not
speculate on how the combination of soft breaking terms, supergravity
corrections and RG improved tree potential lifts this degeneracy to
determine the GUT scale.  We shall assume the necessary vevs and check that
(a) they are consistent with a globally SUSY vacuum and (b)
all states of this sector which are non-singlets under the standard model gauge
group obtain mass of order $M_G$.

The SUSY vacua are specified by :

$$ \langle A_1 \rangle = a_1 \left(\begin{array}{ccccc} 1 &  & & &  \\
& 1 & & & \\ & & 1 & & \\ & & & 0 & \\ & & & & 0  \end{array}\right) \otimes
\eta   $$

$$ \langle A_2 \rangle = a_2 \left(\begin{array}{ccccc} 1 &  & & &  \\
& 1 & & & \\ & & 1 & & \\ & & & -3/2 & \\ & & & & -3/2  \end{array}\right)
\otimes \eta   $$

$$ \langle \tilde{A} \rangle = \tilde{a} \left(\begin{array}{ccccc} 1 &  & &
&\\  & 1 & & & \\ & & 1 & & \\ & & & 1 & \\ & & & & 1  \end{array}\right)
\otimes \eta   $$

$$ \langle S \rangle = s \left(\begin{array}{ccccc} 1 &  & & &  \\
& 1 & & & \\ & & 1 & & \\ & & & -3/2 & \\ & & & & -3/2  \end{array}\right)
\otimes {\bf 1}   $$
where $$ \eta = \left( \begin{array}{cc} 0 & -i \\ i & 0 \end{array} \right)
\; ,  {\bf 1} =  \left( \begin{array}{cc} 1 & 0 \\ 0 & 1 \end{array} \right)
\; .$$
In addition $$  \langle \Psi \rangle = \langle \overline{\Psi} \rangle = V  $$
in the right-handed neutrino direction and
$$ \langle {\cal S}_i \rangle \neq 0 $$
for $i = 1, \cdots, 7$.  The vevs are constrained by the vacuum conditions
$$ {\cal S}_1 + s = 0, \: {\cal S}_2 s + {2 \over 5 }a_1^2 = 0,
\: { V^2 \over 4} + {\cal S}_3 \tilde{a} = 0 $$
$$  {\cal S}_4 a_2 + s \tilde{a} = 0, \: {\cal S}_1 + {\cal S}_5 = 0 .$$

The first term forces $A_1$ into either the $B - L$  or $T_{3R}$ direction but
has a GUT scale barrier to a linear combination of these two directions.  Thus
the $B - L$ vev is natural.  This is made possible by the vev of $S$.  The
third term forces $\tilde{A}$ in the $X$ direction which is consistent with
the SU(5) invariant vevs of $\Psi$ and $\overline{\Psi}$.  The 4th term allows
$A_2$ to have a vev in some linear combination of $Y$ and $T_{3R}$.  Finally,
the terms  $\overline{
\Psi} A_2 \Psi', \overline{\Psi}' A_2 \Psi$  force $A_2$ into the $Y$
direction.  All the other terms in eqn. 4 are necessary to give mass
to all states of this sector which are non-singlets under the standard model
gauge group.

Above the scale $v_{10}$ the gauge coupling is highly non-asymptotically free.
The one loop evolved gauge coupling becomes non-perturbative at about $5
v_{10}$ (depending on both $v_5/v_{10}$ and on values of Yukawa couplings).

\section{Higgs Sector}

In the Higgs sector we distinguish between two cases which differ by the
mechanism for solving the $\mu$ problem.
\begin{enumerate}
\item $\mu$ is generated by a D term coupling of the Higgs doublets to a hidden
sector field $z$ whose F component breaks SUSY at an intermediate scale.
\item $\mu$ is generated by a one loop diagram containing
soft SUSY breaking cubic scalar interactions.  Thus $\mu$ is proportional to
the soft SUSY breaking parameter $A$ \cite{LJHMU}.
\end{enumerate}

In both cases, the reason for $\mu$ having a magnitude which is related to the
scale of supersymmetry breaking is no longer a puzzle. The symmetries of the
theory guarantee that the Higgs doublets are massless at tree level in the
superpotential, and the $\mu$ term arises by introducing supersymmetry
breaking into a higher dimension D operator. The two cases differ as to
the origin of
the D operator. In the first case it is present as a non-renormalizable
operator from the Planck scale, while in the second case it is generated by a
GUT-scale loop.

Higher dimension D operators generate the desired $\mu$ parameter so easily
that its origin should not be considered a problem. Much more serious for
grand unification is the large mass splitting required between the Higgs
doublets and
their color triplet partners. There are several mechanisms which have been
invented to avoid the fine tunings of non-supersymmetric GUTs and of the
Dimopoulos-Georgi supersymmetric GUT \cite{DRW}. In this paper we use the
Dimopoulos-Wilczek mechanism \cite{DW}. In the context of SO(10), this
mechanism requires an adjoint field which points precisely in the $B-L$
direction. Our
theory already contains such a field, $A_1$, with the vev shown in eqn. (2).
This mechanism is a very natural choice in theories where vev
alignments are needed to give fermion mass predictions.

\subsection{Case 1.}
In eqn. 5 we give the Lagrangian for the Higgs sector.
\begin{equation}
L_{Higgs} = [ 10_1  A_1 10_2  +  {\cal S}_7 10_2^2 ]_F
+ {1 \over M}[z^* 10_1^2]_D
\end{equation}
where ${\cal S}_7$ is assumed to get a non-zero vev.
The first two terms (F terms) are necessary to incorporate the
Dimopoulos-Wilczek mechanism for doublet-triplet splitting.
In the doublet sector the vev
$A_1$ vanishes leaving two doublets massless. The massless Higgs doublets lie
solely in the $10_1$.  Thus the SO(10) relation $\lambda_b/\lambda_t = 1$
remains exact.  The last term is a D term and is necessary to solve the $\mu$
problem.  The situation for proton decay is identical to case 2 below and so we
save the discussion until then.

$\bullet$ $\mu$ term

In this case the hidden sector field $z$ is assumed to get a SUSY breaking vev
such that  $F_z \sim \mu M$.

\subsection{Case 2.}
In eqn. 6 we give the superspace potential for the Higgs sector.
\begin{eqnarray}
W_{Higgs} =  &    10_1  A_1 10_2  +  {\cal S}_7 10_2^2  &   \\
& +   {\bar \Psi} {\bar \Psi}' 10_1  +  \Psi \Psi' 10_1   & \nonumber
\end{eqnarray}
The 3rd and 4th terms have been introduced solely to solve the $\mu$ problem.

$\bullet$ Doublet-Triplet Splitting

The mass matrix for the SU(5) ${\bf \overline{5}, 5}$ states in ${\bf 10_1,
10_2, \Psi,  \Psi', \overline{\Psi}, \overline{\Psi}'}$ is given below.
\begin{eqnarray}  &\overline{5}_1  \;\; \overline{5}_2
\;\; \overline{5}_{\Psi}  \;\; \overline{5}_{\Psi'} & \nonumber \\
\begin{array}{c} 5_1 \\ 5_2 \\ 5_{\overline{\Psi}} \\
5_{\overline{\Psi}'} \nonumber \end{array} &
\left( \begin{array}{cccc}  0 & A_1 & 0 & \Psi \\
                            A_1 & {\cal S}_7 & 0 & 0 \\
                            0 &  0 & 0 & A_2 \\
                            \overline{\Psi} & 0 & A_2 & 0 \end{array}  \right)&
\end{eqnarray}

$\bullet$ Higgs doublets

In the doublet sector the vev $A_1$ vanishes leaving two doublets massless.
The massless Higgs doublets, in this case, are a linear combination of the
doublets in 10$_1$  and those in $\Psi$ and $\overline{\Psi}$.  As a result
the boundary condition $\lambda_b/\lambda_t = 1$ is corrected at tree level.
The ratio is now given in terms of a ratio of mixing angles. In principle
this could allow much smaller values of $\tan \beta$ than those predicted when
$\lambda_b/\lambda_t = 1$. However, in practice the predictions for $m_u/m_d$
and for the CP violating parameter $J$ require large $\tan\beta$ \cite{CRW}.
Electroweak symmetry breaking with large $\tan \beta$ can occur with a
moderate fine tune of one part in $\tan \beta$ \cite{EWSB}.

$\bullet$ Proton decay

The leading contribution to proton decay comes from the exchange of superheavy
colored Higgsinos \cite{PDECAY}.
The rate for proton decay in this model is set by the quantity
$(M^t)^{-1}_{11}$ where $M^t$ is the color triplet Higgsino mass matrix. We
find $(M^t)^{-1}_{11}= {{\cal S}_7 \over A_1^2}$, which must be smaller
than ${1 \over \tan\beta M_G}$ in order to be consistent with proton decay
limits. Clearly this relation may be satisfied for ${\cal S}_7$
sufficiently smaller than $M_G$.  Note there are no heavy color triplet
states in this limit.  Proton decay is suppressed because the color triplet
Higgsinos in $10_1$ become Dirac fermions (with mass of order $M_G$), {\em
but they do not mix with each other}.  The ratio ${\cal S}_7/M_G$ cannot be
taken too small however. This is because the doublets in $10_2$ are becoming
lighter than their triplet partners.  This will affect the RG running of the
gauge couplings and is thus constrained by the low energy data.

$\bullet$ $\mu$ term

In this case $\mu$ is generated at the one loop level, for example by the
diagram of fig. 2.  We find
$\mu \sim {\lambda^4 \over 16 \pi^2} A$.   Thus $\mu$ is naturally smaller
than the scale of SUSY breaking.  This is a nice feature since it is
consistent with a large $\tan \beta$ solution.

\section{Neutrino Masses}

Although right-handed neutrinos are included in the $16_i$, $i= 1,2,3$ this
does not mean that the three electroweak doublet (left-handed) neutrinos are
massive.

An $R$ symmetry, discussed in the next section, forbids operators of the form
$16_i \overline{\Psi} 16_j \overline{\Psi}$ which could generate Majorana
masses for the right-handed
neutrinos. To avoid Dirac neutrinos with weak-scale masses, extra fields must
be introduced which couple to $\nu_R$ at the renormalizable level.
The form of these interactions is very model dependent.
In the simplest case the three left-handed
neutrinos are massless.  This is case 1. In case 2 we show how to generate
mass for the left-handed neutrinos.

\subsection{Case 1. - Massless Neutrinos}

In this case we introduce  three  SO(10) singlets fields  $N_i,  i = 1,2,3$;
one for each family. The superspace potential for neutrinos is given by

\begin{equation}
W_{neutrinos}  =     \sum_{i = 1}^3 (16_i \overline{\Psi}) N_i
\end{equation}
When $\overline{\Psi}$ gets a vev the right-handed neutrinos mix with the
singlet states to get Dirac masses of order $V$.  The left-handed neutrinos
remain massless.

\subsection{Case 2. - Massive Neutrinos}

In this case there are several apriori acceptable possibilities.  The
different choices are described in terms of an effective symmetric mass
matrix $M_{ij}= M_{ji}$ for the states $N_i, i=1,2,3$.  For every zero
eigenvalue of $M_{ij}$ there is a corresponding massless neutrino.  Thus with
only one term, for example $M_{23}$, the $\mu$ and $\tau$ neutrinos would be
massive and the electron neutrino would be massless.  There would nevertheless
be mixing among all three.

Choosing a set of operators will affect the symmetries of the theory.  We can
only accept operators which preserve enough symmetry to keep the theory
natural. One example of an allowed neutrino mass sector is given by the
following superspace potential.
In this case we have effectively $M_{22}$ and $M_{13}$ non-zero.

\begin{eqnarray}
W_{neutrinos}  = &\sum_{i = 1}^3 (16_i \overline{\Psi}) N_i  &   \\
& +  N_2^2 {\cal S}_M +  N_1 N'_1 {\cal S}_M & \nonumber \\
 &+ N_3 N'_3 {\cal S}_M +  N'_1 N'_3 {\cal S}  & \nonumber
\end{eqnarray}
where  ${\cal S} = {\cal S}_1$ or ${\cal S} = {\cal S}_3$.  Upon integrating
out ${\cal S}_M$ we could obtain an equivalent description in terms of the
higher dimension operators
\begin{equation}
W_{neutrinos}  = {1 \over {\cal S}_M} (16_2 \overline{\Psi})^2  +  {1 \over
{\cal S}_M^2} (16_1 \overline{\Psi})(16_3 \overline{\Psi}) {\cal S}
\end{equation}
Inserting $\overline{\Psi}$ vevs yields masses for the $\nu_R$ in the $16_i$,
suppressed by various powers of $v_{10}/M$ and $M_G/M$. These masses take part
in the see-saw mechanism giving small masses to the left-handed neutrinos.
The suppression factors for the $\nu_R$ masses leads to enhancements of the
masses of the light neutrinos, which can have phenomenologically important
consequences. Since there are two operator coefficients in equation 9, the 3
masses and 3 real mixing angles of the light neutrinos are given in terms of
only 3 unknown parameters (two magnitudes and a phase).

\section{Symmetries}

The theory of this paper is the most general consistent with several
symmetries.

$\bullet$  Continuous R Symmetry --- Since the theory is scale invariant it
has a continuous R symmetry in which all superfields have charge 1 and the
superspace potential has charge 3. With this symmetry we need only consider
dimension 4 operators in the superspace potential below $M$. This $R$ symmetry
is a U(1) Peccei-Quinn symmetry. It solves the strong CP problem and leads to
an invisible axion \cite{NR}.

$\bullet$  Family Reflection Symmetry\footnote{See Dimopoulos and Georgi
\cite{DRW}.}  --  This symmetry eliminates dimension 4 baryon number
violating operators and is equivalent to R parity.  Thus the
LSP is stable.  Under this symmetry all states in the set \{$16_i, N_i,\: i =
1,2,3; \: N'_1, N'_3;  \: \Psi_a,$ $ \overline{\Psi}_a, \: a = 1, \cdots , 9$
\}
are odd and all other states are even.

$\bullet$  Z(4) R Symmetry  --  This symmetry guarantees that only $10_1$
couples to fermions.  Under this symmetry all primed fields, ${\cal S}_6,
{\cal S}_7$ and $10_2$ are odd;  the set \{ $16_i, \: i=1,2,3; \: \Psi_a,
\overline{\Psi}_a, \: a= 1, \cdots, 9 $ \}  transform by $ i\times
\{ \cdots \}$;  all other fields are even.

$\bullet$  U(1) Flavor Symmetries. These symmetries guarantee that the
dimension 4 operators included in equations (3 - 8)  are ``natural" in the
sense that no other dimension 4 operators are consistent with the family
reflection symmetry, the Z(4) R symmetry and the U(1) symmetries.
These U(1) symmetries are the only symmetries of the theory which distinguish
between the three generations, $16_i$. They are
largely responsible for determining which flavor operators are generated;
hence, we call them flavor symmetries. The more terms in the superspace
potential the fewer continuous symmetries exist. For example, consider the
Higgs sector as in case 2 (eqn. 6) and the neutrino sector of eqn. (8).
This theory has 49 fields and 47 superpotential interactions, hence there
are two U(1) flavor symmetries. We have checked that these two symmetries
rule out the addition of any new terms to W.
The other Higgs sectors and neutrino sectors which we have discussed have
fewer interactions, and correspondingly more U(1) symmetries.

The above symmetries are sufficient to limit the flavor interactions of the
superpotential to those which lead to the desired fermion mass predictions.
Furthermore, since all non-renormalizable F terms are forbidden, the
predictions can be computed exactly in terms of the renormalizable
coefficients. In general it is possible that higher dimension D operators,
allowed by the above symmetries, could lead to corrections to the flavor
predictions. However, if the supersymmetry breaking occurs in a hidden sector
of a supergravity theory, flavor changing phenomenology leads to considerable
constraints on the form of the higher dimension D operators.
This phenomenology requires scalar masses
and trilinear interactions to be close to universal,
suggesting that the D terms have a trivial flavor
structure, as would occur if they possessed a $U(N)$ invariance \cite{HLW}.
In this case the higher dimension D operators do not affect the fermion mass
predictions. A further consequence of the $U(N)$ invariance of the K\"ahler
potential is that the operator $[z^* \; 10_1^2]_D$ of equation (5), which was a
possible source of the $\mu$ term, is absent. Hence, the $\mu$ term should be
generated radiatively as in section (4.2). \footnote{We can envisage other
schemes for the D terms. In the limit that the gaugino mass is much larger than
the scalar mass and A terms at $M_P$, the flavor changing problem is alleviated
and there is no need for universal scalar masses and A terms. In this case
the D terms could give corrections to the fermion mass predictions, suppressed
compared to the leading terms by powers of $v_{10}/M_P$.
Also the $\mu$ term could occur via the operator $[z^* \; 10_1^2]_D$.}

In a supersymmetric theory, it may be questioned whether there is any need to
guarantee the form of the desired interactions by symmetries. It has been
argued that string theories can lead to low energy effective theories with
interactions which are not the most general allowed by the low energy
symmetries. This makes the task of model building easier. It is probable that
a much simpler grand unifed SO(10) theory can be written down if one adopts
this viewpoint. However, in a sense it makes model building too easy. For
example, as a flavor sector
one can simply write down the four flavor operators of
equation 1. There is no symmetry to forbid all other operators, but no such
symmetry is needed. Further operators can be added at will. While model
building is much easier in this approach, it leads to much less understanding:
the origin of flavor is hidden in physics above the string compactification
scale. By contrast, in this paper flavor physics is determined by a set of
global U(1) symmetries beneath the string scale \footnote{Another possibility,
which has been explored recently, is that the pattern of fermion masses arises
from a U(1) flavor symmetry which is gauged \cite{IR}.}

\section{Conclusions}

We have presented a complete SO(10) SUSY GUT incorporating the fermion mass
predictions of model 6  of ref.\cite{ADHRS}.  We have addressed the issues of
doublet-triplet splitting, proton decay, $\mu$ problem and fermion masses and
mixing angles.  In the SUSY limit the superspace potential has many flat
directions, one of which appears consistent with low energy data. In
particular, in this desired vacuum the GUT vevs align precisely in the
directions necessary to yield light Higgs doublets and several fermion mass
predictions.
The lifting of the degeneracy of these flat directions has not been studied,
and will lead to interesting constraints on the form of the supersymmetry
breaking terms. The flat directions imply the existence of light
supermultiplets which are singlets under the standard model gauge group.
Moreover since the continuous U(1) symmetries are spontaneously broken at $
M_G$ the theory also includes an invisible axion and
possibly several massless Goldstone bosons.  These may be problematical
in a cosmological context and also require further study.

In this theory we can now calculate corrections to zeroth order fermion mass
and mixing angle predictions\cite{ADHRS}. Threshold corrections at the GUT
scale include corrections to (3) coming from integrating out the heavy
$16 - \overline{16}$ pairs (only the zeroth order result in a power series in
ratios of scales has been included
in the analysis of ref. \cite{ADHRS}).  There are also radiative corrections
to gauge and Yukawa parameters; for example, the radiative corrections to the
gauge couplings can be significant in the limit ${\cal S}_7 \ll M_G$ and thus
the low energy measurement of $\alpha, sin^2\theta_W$ and $\alpha_s(M_Z)$
could constrain the proton decay rate.

Finally, several groups are now working to obtain SO(10) SUSY GUTs from string
theories \cite{CCL}.  We hope that theories like ours can be used as a guide
to find a realistic fundamental theory of nature.

{\bf Acknowledgements}

We thank R. Barbieri and A. Strumia who participated in an earlier version of
this work.

\newpage

{\bf Figure Captions}

1. Diagrams which generate the operators of equation 1.

2. A diagram which generates a one loop contribution to $\mu$ by integrating
   out superheavy fields.

\end{document}